\documentclass[12pt]{article}
\usepackage{amsmath}
\usepackage{amssymb}
\usepackage{authblk}
\usepackage{amsthm}
\usepackage[mathscr]{eucal}
\usepackage[usenames]{color}
\usepackage[utf8]{inputenc}
\usepackage{url}

\usepackage[
bookmarks=true,
backref=true,
colorlinks=true,
linkcolor=red,
citecolor=blue,
urlcolor=green]{hyperref}

\numberwithin{equation}{section}
\numberwithin{theorem}{section}
\numberwithin{lemma}{section}
\numberwithin{proposition}{section}
\numberwithin{corollary}{section}
\numberwithin{remark}{section}
\numberwithin{definition}{section}

\newcommand{\gen}[1]{\partial_{#1}}

\newcommand{\curl}[1]{ \{#1\} }
\newcommand{\lie}{\mathfrak g}

\definecolor{darkolivegreen}{rgb}{0.333333, 0.419608, 0.1843140}

 \DeclareMathOperator{\conf}{c}
 
\DeclareMathOperator{\Poinc}{P}
\DeclareMathOperator{\Ort}{O}
\DeclareMathOperator{\Conf}{C}
\DeclareMathOperator{\euclid}{e}
\DeclareMathOperator{\G}{G}
\DeclareMathOperator{\Div}{Div}

\newcommand{\dt}{\partial_t}

\begin{document}

\title{Noether symmetries and conservation laws of a class of time-dependent multidimensional nonlinear wave equations}

\author[1]{F. G\"ung\"or\thanks{e-mail: gungorf@itu.edu.tr}}
\author[1]{C. \"Ozemir\thanks{e-mail: ozemir@itu.edu.tr}}

\affil[1]{Department of Mathematics, Faculty of Science and Letters, Istanbul Technical University, 34469 Istanbul, Türkiye}


\date{}

\maketitle

\begin{abstract}
Conservation laws of a class of time-dependent damped nonlinear multidimensional wave equations   are derived by Noether's theorem. For arbitrary nonzero damping coefficient and nonlinear interaction term, its infinitesimal variational symmetries span a Euclidean algebra $\euclid(n)$ of space translations and rotations. They produce conservation of linear and angular momentums. For some specific forms of these two terms symmetry algebra is enlarged to a subalgebra of the conformal algebra $\conf(1,n)$ and in this case more interesting conservation laws are found.

\end{abstract}


\section{Introduction}
The aim of this paper is to study symmetries and conservation laws of the  nonlinear wave equations
\begin{equation}\label{main}
 \mathcal{E}=   \Box u+a(t)u_t+f(u)=0, \quad t>0, \quad f''\neq 0
\end{equation}
 where $\Box =\dt^2-\Delta$ is the $(n+1)$-dimensional wave operator, $u(t,x)$ is the wave function with $(t,x)\in \mathbb{R}^{n+1}$, $n\geq 2$ and $f(u)$ is an arbitrary function.

Equation \eqref{main} is the Euler-Lagrange equation $E_u(L)=-\mu\mathcal{E}$ with the Lagrangian
\begin{equation}\label{L}
 L=\mu(t)L_0, \quad L_0=\frac{1}{2}\big(u_t^2-|\nabla_{x} u|^2\big)-F(u), \quad F(u)=\int^{u} f(z)dz
\end{equation}
where $\mu>0$ satisfies $\dot \mu=a(t)\mu$, $\nabla_{x}$ and  $E_u$ are the spatial gradient and the Euler-lagrange operators, respectively. Quite recently, the same class in the one-dimensional case has been investigated in \cite{AncoMarquezGarridoGandarias2024}. The authors of this paper use a multiplier approach to find conservation laws. Here, in the multi-dimensional case, we prefer to take the variational symmetry approach with the same goal in mind. Very recently, Noether symmetries and first integrals for the general class of Lagrangians
$L = \mu(t)L_0$, which models motion with general linear damping in a Riemannian space have been derived in Ref. \cite{Tsamparlis2026}.

Obviously, $L_0$ in \eqref{L} is the Lagrangian of the undamped nonlinear wave equation obtained when $a(t)\equiv 0$,
 \begin{equation}\label{undamped}
\Box u+f(u)=0.
 \end{equation}
In a recent paper \cite{GuengoerOezemir2026}, we have analysed Lie point symmetries of  more general classes of equations
\begin{subequations}
    \begin{eqnarray}   \Box u  &=& G(x,u,\nabla u),  \label{GO1}\\
  \Box u  &=& G(x,u),  \label{GO2}
    \end{eqnarray}
\end{subequations}
where $x=(t,x)=(t,x_1,\ldots,x_n)$ and $\nabla=(\partial_0,\nabla_x)=(\partial_t,\partial_{x_1},\ldots,\partial_{x_n})$ is the gradient operator in $\mathbb{R}^{1,n}$.
Throughout this paper, unless otherwise stated we will sum over repeated indices. Eq. \eqref{main} belongs to the first class \eqref{GO1} with the special choice
$$G(x,u,\nabla u)=-a(t)u_t-f(u), \quad f''\ne 0,$$
which is a nonlinear extension of the linear wave equation
\begin{equation}\label{linearwave}
    \Box u=0.
\end{equation}
A well-known fact about \eqref{linearwave} is that it is invariant under the conformal symmetry group $\mathsf{C}(1,n)$, isomorphic to the pseudo-orthogonal group $\mathsf{O}(n+1,2)$.

For the undamped wave equation \eqref{undamped}, $a(t)=0$,  the energy
\begin{equation}\label{energy}
E(u)=\int_{\mathbb{R}^n}\Big[\frac{1}{2}\big(u_t^2+|\nabla_{x} u|^2\big)+F(u)\Big]dx
\end{equation}
is a constant of the motion (conserved quantity). On the other hand, when $a(t)\ne 0$,
we have the relation
\begin{equation}\label{energy-decr}
\frac{dE}{dt}=-a(t)\int_{\mathbb{R}^n}u_t^2dx \leq 0
\end{equation}
and the energy is monotonically decreasing for $a>0$. Another conserved quantity is angular momentum (see below formula \eqref{angular-comp}).

A number of interesting conservation laws of \eqref{linearwave} were derived in \cite{Olver1993} (see examples 4.15 and 4.36) based on Noether's  theorem.  We shall make use of the same theorem to derive conservation laws of the class of \eqref{main} and its subclasses allowing maximal symmetry algebra.

\section{Symmetries and Conservation Laws}
A general element of a Lie symmetry algebra $\lie$ will be a vector field of the form
\begin{equation}\label{vf}
  \mathbf{v}=\sum_{a=0}^{n}\xi_{a}\gen{x_{a}}+\eta \gen u,
\end{equation}
where the coefficients $\xi_{a}$ and $\eta$ are functions of $(t,x,u)$ with the identification $x_0=t$.

The Lie symmetry algebra  $\lie=\conf(1,n)$ of \eqref{linearwave} has dimension $\dim \lie=(n+2)(n+3)/2$ and is spanned by the vector fields
\begin{equation}\label{killing-vf}
  \begin{split}
P_{a}&=\partial_{a},\\
J_{ab}&=x_{a}\partial_{b}-x_{b}\partial_{a},\\
D&=x^{a}\partial_{a} + \frac{(1-n)}{2}u\partial_u,\\
C_{a}&=2x_{a}D-(x_{\alpha}x^{\alpha})\partial_{a},
\end{split}
\end{equation}
where $a, b= 0, 1, \dots, n$.

Preferably, $C_a$ can be written as
\begin{equation}\label{conf-gen}
\begin{split}
C_0&=(t^2+r^2)\dt+2tx_l\partial_l+(1-n)tu\partial_u,\\
C_k&=2tx_k\partial_t+2x_k x_l \partial_l+(t^2-r^2)\partial_k+(1-n) x_k u \partial_u,
\end{split}
\end{equation}
where $r^2=|x|^2=\sum_{i=1}^{n}x_i^2$, $k, l=  1, \dots, n$.

For arbitrary $a(t)$ and $f(u)$, the symmetry algebra of \eqref{main} is the Euclidean algebra $\euclid(n)$ of translations and rotations of $\mathbb{R}^n$. Its dimension is $n(n+1)/2$ and infinitesimal symmetries are
\begin{equation}\label{sym-damped}
P_{k}=\partial_{k}, \quad
J_{kl}=x_{k}\partial_{l}-x_{l}\partial_{k}, \quad k<l, \quad k, l=  1, \dots, n.
\end{equation}
We note that  the (equivalence) transformation $u\to u+u_0$, $u\to \lambda u$  takes  Eq. \eqref{main} to the same form with $f(u)$ changed.
If we require invariance under the dilational symmetry
\begin{equation}\label{dilat}
  D=x^{\sigma}\partial_{\sigma} + d u\partial_u=t\gen t+x_k \gen k+d u\gen u,
\end{equation}
then we find that for
\begin{equation}\label{a-and-f}
  a(t)=\frac{m}{t}, \quad f(u)=f_0 u^p, \quad p\ne 1, \quad d=\frac{2}{1-p},
\end{equation}
where $m$ and $f_0$ are integration constants.
On the other hand, for $m\in \mathbb{R}$, dilatational invariance under
\begin{equation}\label{D-2}
  \tilde{D}=t\gen t+x_k \gen {k} -\frac{2}{\alpha} \partial_u,  \quad \alpha\ne 0
\end{equation} leads to
the exponential nonlinearity
\begin{equation}\label{f-exp}
  f(u)=f_0 e^{\alpha u}.
\end{equation}

Further special conformal extension of $D$ is possible
\begin{equation}\label{Kk}
  C_k=2tx_k\partial_t+2x_k x_l \partial_l+(t^2-r^2)\partial_k+q x_k u \partial_u
\end{equation}
when  $p$ is restricted to be
\begin{equation}\label{k}
 p=\frac{n+3+m}{n-1+m},  \quad m\ne 1-n.
\end{equation}
with $q=1-n-m=d/2$.
We hereby have corrected a misprint concerning the conformal factor $q$ in our paper \cite{GuengoerOezemir2026}.

So we will be dealing with the damped PDE (partial differential equation)
\begin{equation}\label{damped-pde}
  \Box u+\frac{m}{t}u_t+f_0 u^p=0, \quad p=\frac{n+3+m}{n-1+m},
\end{equation}
or
\begin{equation}\label{damped-pde-exp}
   \Box u+\frac{m}{t}u_t+f_0 e^{\alpha u}=0, \quad \alpha\ne 0.
\end{equation}

In the exponential case \eqref{damped-pde-exp}, a conformal extension occurs when $m=1-n$. In this case $C_k$ are replaced by
\begin{equation}\label{Kk-exp}
  C_k=2tx_k\partial_t+2x_k x_l \partial_l+(t^2-r^2)\partial_k-\frac{4}{\alpha} x_k  \partial_u, \quad \alpha\ne 0, \quad k=1,2,\ldots, n.
\end{equation}
In both cases the Lie symmetry algebra has dimension of $(n+1)(n+2)/2$ and $f_0$ can be set to $f_0=\pm 1$ using the above-mentioned equivalence transformation.

Some special cases of $p$ are in order. If $m=-(n+1)$, then $p=-1$ ($d=1$) and we have $f(u)=f_0 u^{-1}$ (the potential is logarithmic function). For $m=-(n+3)$, $p=0$ which is not allowed due to the assumption $f''\ne 0$.
Finally in the undamped case $m=0$, Eq. \eqref{damped-pde} reduces to the well-known conformally-invariant nonlinear wave equation with Lie symmetry algebra (factoring out the symmetry representing the linear superposition principle) shared with the homogenous linear wave equation \eqref{linearwave}. For an alternative derivation of \eqref{damped-pde} in the general case $n\geq 2$, see Refs. \cite{GuengoerOezemir2026, FushchichShtelenSerov1993} or  \cite{Guengoer1998, Guengoer2019} for $n=2$. In the undamped case $m=0$ with power non-linearity $f(u)=\pm u^p$ where $u=u(t,r)=u(t,|x|)$, a study of conservation laws was given in \cite{AncoIvanova2006}. The interested reader can also find basic definitions, ideas and interesting results applied to a lot of PDEs in \cite{Anco2016, Anco2017, AncoGandariasRecio2018}.

Our primary interest is to derive conservation laws of the damped nonlinear wave equation \eqref{damped-pde} utilizing variational structure  of the equation. This will be done by means of possible variational symmetries.

A (local) conservation law is an expression of the form
\begin{equation}\label{conser-exp}
  \Div \tilde{I}=D_t I_0+\Div I=0, \quad \tilde{I}=(I_0,I)=(I_0,I_1,\ldots, I_n),
\end{equation}
which holds for all solutions of the PDE \eqref{main}. Here $I_0$ is the conserved density, $I=(I_1,\ldots, I_n)$ the  associated (spatial) flux all of which are functions of $(t,x,u)$ and the derivatives of $u$ with respect to the independent variables $(t,x)\in \mathbb{R}^{n+1}$.
Here $\Div I$ is defined to be the spatial total derivative of the $n$-tuple $I$ with respect to $(t,x)$
$$\Div I=\sum_{j=1}^n D_{j} I_{j},  $$
where $D_{j}$ is the total differential operator defined by
$$D_{j}=\gen{j}+u_{j}\gen{u}+ u_{jk}\gen{u_{k}}+\ldots $$
For a more detailed discussion on this theme and related issues  we refer to \cite{Olver1993, Olver1995, HydonKing2025}.

The following is the infinitesimal criterion of invariance for a first order Lagrangian $L$:
\begin{equation}\label{inv-L}
\mathrm{pr}^{(1)} \mathbf{v} (L)+L\Div{\xi}=\Div  \tilde{B}=D_t B_0+\Div  B,
\end{equation}
where $\xi=(\xi_0,\xi_1,\ldots,\xi_n)$ and $B$ is a $(n+1)$-tuple of differential function  in the jet space $J(t,x,u,u_{a})$. The evolutionary form $\mathbf{v}_{Q}=Q\gen u$ of the vector field \eqref{vf} permits us to express \eqref{inv-L}
\begin{equation}\label{inv-L-evo}
  \mathrm{pr}^{(1)} \mathbf{v}_{Q} (L)=\Div B,
\end{equation}
where $Q(t,x,u,u_a)$ is the characteristic function of the vector field $\mathbf{v}$ of \eqref{vf} and $B$ is another $(n+1)$-tuple differential function.
In terms of $Q(t,x,u,u_a)$ of the variational symmetries we can express the conservation laws in characteristic form
\begin{equation}\label{charac-form}
\Div I =Q.E_u(L)=-\mu Q.\mathcal{E}=0,
\end{equation}
where $\mathcal{E}=0$ is the wave equation defined by \eqref{main}. For a first order Lagrangian, the constants of motion or first integrals are given either by integrating by parts \eqref{charac-form} or explicitly by the formula
\begin{equation}\label{cons-law-comp}
I_a=\xi_a L+Q \frac{\partial L}{\partial u_a}=\mu(t)\left[\xi_a L_0+Q \frac{\partial L_0}{\partial u_a}\right], \quad a=0,1,\ldots, n.
\end{equation}

As was also observed in Ref. \cite{AncoMarquezGarridoGandarias2024}, for the Lagrangian $L=\mu L_0$ given by \eqref{L}, the action of $L$ on the first prolongation of $\mathbf{v}_{Q}$ is
\begin{equation}\label{pr-vf}
 \mathrm{pr}^{(1)} \mathbf{v}_{Q} (L)=-\mu Q \mathcal{E}+D_t(\mu u_t Q)-\Div (\mu Q \nabla_x u)
\end{equation}
and
the invariance condition \eqref{inv-L-evo} takes the form
\begin{equation}\label{multip-form}
   \mathrm{pr}^{(1)} \mathbf{v}_{Q} (L)=-\mu Q \mathcal{E}+D_t(B_0-\mu u_t Q)+\Div (B+\mu Q \nabla_x u).
\end{equation}
This shows that if $\mathbf{v}_{Q}$ generates a variational symmetry then $\tilde{Q}=\mu Q$ is a multiplier for the corresponding conservation laws, namely,
 \begin{equation}\label{first-int}
  \mu Q. \mathcal{E}=\Div  \tilde{I}=D_t I_0+\Div I=0,
 \end{equation}
and $E_u(\mu Q. \mathcal{E})$ vanishes identically.

Integrating the conservation laws \eqref{first-int} over a bounded domain $\Omega\subset \mathbb{R}^n$, using divergence theorem under the assumption that the flux term (an $n$-tuple of functions of $(t,x,u)$ and  first order derivatives of $u$) $I$ tends to zero as $x$ approaches the boundary $\partial \Omega$, or if $\Omega=\mathbb{R}^n$ as $|x|\to \infty$ we can construct a constant of the motion of \eqref{main}  in the integral form
\begin{equation}\label{const-motion}
  \int_{\Omega} I_0 \; dx=\text{const.}
\end{equation}

\subsection{Conservation laws for \eqref{main}}

We now turn to the derivation of conservation laws for the original class \eqref{main} for any function $a(t)$, and start with Noether symmetries  corresponding to the Euclidean algebra $\euclid(n)$. The characteristic function for $P_k$ is $Q=u_k$ (up to sign), $k=1,2,\ldots, n$ and integration by parts of the equation
\begin{equation}\label{cons-Pk}
  \Div \tilde{I}=\mu u_k \mathcal{E}=D_t(\mu u_k u_t)+\mu D_j\left[-u_k u_j+(\frac{1}{2}\left(u_t^2-\left|\nabla_x u \right|^2-2F(u)\right)\delta_{kj}\right].
\end{equation}
So we have the conserved density and the current components of the linear momentum conservation
\begin{equation}\label{cons-Pk-comp}
I_0^{k}=\mu u_k u_t,  \quad I_j^{k}=\mu(-u_k u_j+ \delta_{kj} L_0), \quad \mu=\exp{\int a(t)dt}, \quad j,k=1,2,\ldots, n.
\end{equation}

A similar manipulation gives the angular momentum conservation for the characteristic function $Q_{kl}=x_k u_l-x_l u_k$, $k<l$ of the geometrical rotations $J_{kl}$
\begin{equation}\label{angular}
  \mu (x_k u_l-x_l u_k) \mathcal{E}=D_t(\mu Q u_t)+\mu D_j[x_l \Phi_{jk}-x_k \Phi_{jl}+(\delta_{jl}x_k-\delta_{kj} x_l)F],
\end{equation}
where the symmetric function $\Phi_{jk}$ is defined as
$$\Phi_{jk}=u_j u_k+\frac{1}{2}(u_t^2-\left|\nabla_x u \right|^2) \delta_{jk}.$$
From this we see that the components of the conservation laws are in the form
\begin{equation}\label{angular-comp}
  I_0^{kl}=\mu Q_{kl} u_t,  \quad I_j^{kl}=\mu [-Q_{kl}u_j+(\delta_{jl}x_k-\delta_{kj} x_l)L_0], \quad j,k,l=1,2,\ldots, n.
\end{equation}

Now we wish to determine if the subclasses \eqref{damped-pde} and \eqref{damped-pde-exp}  admit variational symmetries. Our main tool  is the infinitesimal criterion \eqref{inv-L} (necessary and sufficient for a connected group of transformations to  be a symmetry  of a variational problem). Let us start with the dilational generator
\begin{equation}\label{D-d}
  D=t\gen t+x_k \gen  k+d u\gen u, \quad d\ne 0
\end{equation}
and compute the left side of \eqref{inv-L} as
$$\mathrm{pr}^{(1)} D (L)+(n+1)L=(t \dot{\mu}+(n+1)\mu+2\mu(d-1))L_0-\mu(d u F'-2(d-1)F)$$
using
$$\Div \xi=n+1, \quad \mathrm{pr}^{(1)} D=D+(d-1)u_i \gen {u_i}. $$
The coefficients of $L_0$ and $\mu\ne 0$ in this relation imply
$$t \dot{\mu}+(n+1)\mu+2\mu(d-1)=\mu[t a(t)+(n+1)+2(d-1)]=0,  \quad d u F'-2(d-1)F=0.$$
The first relation suggests that we must have $t a=m=\rm constant$ and $\mu=t^m$ so we find the dilational factor $d=(1-m-n)/2$. The second one in terms of $f$ is
$$d u f'-(d-2)f=0,$$ which has the solution
\begin{equation}\label{max-power-f}
  f=f_0 u^p, \quad p=\frac{d-2}{d}=\frac{n+3+m}{n-1+m},
\end{equation}
exactly as in the form \eqref{damped-pde}. Consequently, this result confirms that $D$ with $d=(1-m-n)/2$ is a variational symmetry for arbitrary $m$ and $f(u)$ as in \eqref{max-power-f}. Taking
$$Q=d u-t u_t-x_j u_j, \quad L_0=\frac{1}{2}\left(u_t^2-\left|\nabla_x u \right|^2\right)-\frac{f_0}{p+1}u^{p+1}, \quad p+1=\frac{2(n+1+m)}{n-1+m}$$ from \eqref{cons-law-comp} it follows the following conserved currents
\begin{equation}\label{dil-currents}
  I_0=t^m[t L_0+Q u_t], \quad I_j=t^m[x_j L_0+Q u_j].
\end{equation}

The same calculation for \eqref{D-2} gives $m=1-n$ which is a dimension-dependent constraint. Hence, only for these values of $m$,   conservation laws  can be obtained from  \eqref{cons-law-comp}.

We can repeat the same steps for the conformal symmetry generators $C_k$ with the conformal factor $q$, that is ab initio left free.
We take into account the relations
$$\Div \xi=2(n+1)x_k, \quad \mathrm{pr}^{(1)} C_k=C_k+\tau_{k,0}\gen {u_t}+\tau_{k,l} \gen {u_l},$$
where the coefficients  of the first prolongation are functions of $(t,x,u,u_a)$ defined by
$$\tau_{k,l}=D_{l}Q_k+u_{jl}\xi_{j}=D_{l} \eta_k -u_{j}D_{l} \xi_j, \quad j,k=1,2,\ldots,n.$$
If we take the coefficients $\xi$ and $\eta$ of  $C_k$  as
$$\xi_{k,0}=2t x_k, \quad \xi_{k,l}=2x_k x_l+(t^2-r^2)\delta_{kl},  \quad \eta_k=q x_k u$$
and use the relations
$$2tx_k\dot{\mu}=2m \mu x_k, \quad u_t\tau_{k,0}=(q-2)x_k u_t^2-2t u_t u_k, \quad -u_l \tau_{k,l}=-quu_k-(q-2)x_k\left|\nabla_x u \right|^2+2tu_t u_k$$
we obtain
\begin{multline}\label{Ckvar}
  \mathrm{pr}^{(1)} C_k(L)+2(n+1)x_k L= \\
  -q \mu u u_k+\mu(q+m+n-1)x_k[u_t^2-\left|\nabla_x u \right|^2]-\mu[q u F'+2(m+n+1)F]x_k.
\end{multline}
The left side of this expression is a divergence $\Div B_k=D_a B_{k,a}$ with components
$$  B_{k,0}=0 ,\quad  B_{k,j}=-\frac{q}{2}t^m u^2 \delta_{kj}$$
if we choose
$$q=1-m-n, \quad qu f'+[q+2(m+n+1)]f=0,  \quad f=F'.$$
The second relation shows that $f=f_0 u^p$, $p=(n+3+m)/(n-1+m)$  and the conformal transformations \eqref{Kk} are also variational (more precisely divergence) symmetries. The conserved currents  of the corresponding conservation law  from formula \eqref{cons-law-comp} with $I_a$ replaced by $I_{a}-B_{a}$ for each $a\in\curl{1,2,\ldots,n}$ are
\begin{equation}\label{consv-law-conf}
I_{k,0}=t^m(2t x_k L_0+Q_k u_t), \quad   I_{k,j}=t^m[L_0 \xi_{k,j}-Q_k u_j+\frac{q}{2}u^2 \delta_{kj}], \quad j,k=1,2,\ldots, n.
\end{equation}
In \eqref{cons-law-comp}, $\xi_{k,j}$ denotes the horizontal coefficients   of the $k$-th components of  $C_k$ and $Q_k$  is the characteristic function of $C_k$ in \eqref{Kk}.

Finally, we check the formula \eqref{inv-L} for $C_k$ given in \eqref{Kk-exp}, and generic $f(u)$ and find
\begin{equation}\label{check-inv-conf-exp}
  \mathrm{pr}^{(1)} C_k(L)+2(n+1)x_k L=\frac{4}{\alpha} \mu u_k+\mu\alpha(m+n-1)x_k [u_t^2-\left|\nabla_x u \right|^2]+\frac{2\mu}{\alpha}[(m+n+1)F-2F'].
\end{equation}
When $m=1-n$ and $F(u)=F_0 e^{\alpha u}$, the right side of \eqref{check-inv-conf-exp} is a divergence  in the form
$$\Div B_k=D_t B_{k,0}+D_j B_{k,j},  \quad B_{k,0}=0,  \quad B_{k,j}=\frac{4}{\alpha}t^m u \delta_{kj}.$$
The corresponding dimension-dependent conserved currents  are
\begin{equation}\label{comp-exp}
 I_{k,0}=t^{1-n}(2 x_k L_0-Q_k u_k), \quad   I_{k,j}=t^{1-n}[L_0 \xi_{k,j}-Q_k u_j-\frac{4}{\alpha}u \delta_{kj}], \quad j,k=1,2,\ldots, n.
\end{equation}

For any global solution to the damped wave equation \eqref{main} decaying sufficiently rapidly as $r=|x|\to \infty$, the spatial integrals of the  conserved densities $I_0$ found above provide us with constants of the motion of  \eqref{main}.

We recall that in the undamped case $m=0$ \eqref{undamped}, for any $f(u)$, Eq. is invariant under the  Poincar\'{e} group $\Poinc(1,n)$ which is a semidirect product of the Lorentz group with the group of space-time translations: $\G=\Poinc(1,n)=\mathbb{R}^{1,n}\rtimes \Ort(1,n)$. This group of  transformations are variational symmetries and has dimension $\dim {\G}=(n+1)(n+2)/2$. All conservation laws associated to them can be found by \eqref{cons-law-comp}.

The symmetry group for the special power linearity $f(u)=f_0u^p$, $p=(n+3)/(n-1)$ is even larger, the conformal group $\Conf(1,n)$. In this case, there are $n+2$ more conservation laws obtainable from \eqref{cons-law-comp}. One of them is of course the conservation of energy $E$
\begin{equation}\label{E}
  E=[\frac{1}{2}\big(u_t^2+|\nabla_{x} u|^2\big)+F(u)\Big], \quad F(u)=\frac{f_0(n-1)}{2(n+1)}u^{2(n+1)/(n-1)}
\end{equation}
with the multiplier $Q=u_t$. The flux (current) term is $I_j=u_t u_j$.

The remaining symmetry generators consisting of one more conformal  and all Lorentz generators
\begin{equation}\label{C0,Kk}
  C_0=(t^2+r^2)\dt+2tx_l\partial_l+(1-n)tu\partial_u,  \quad K_k=J_{k0}=x_k \gen t+t\gen {k}, \quad k=1,2,\ldots,n
\end{equation}
are also variational.
Indeed, for $C_0$ the infinitesimal variational criterion is satisfied with the gauge function $B_0=-((1-n)/2) u^2$. The density and flux components of the conservation law has the form
\begin{equation}\label{C0-CL}
  I_0=(t^2+r^2)L_0+Q_0u_t+\frac{n-1}{2}u^2, \quad I_j=2tx_j L_0-Q_0 u_j,
\end{equation}
where
$$Q_0=(1-n)tu-(t^2+r^2)u_t-2tx_ku_k$$ is the characteristic function of $C_0$.

The conservation laws for the Lorentz boosts are given by
\begin{equation}\label{Lorentz-CL}
  D_t[x_kE+tu_tu_k]+D_j[tE+x_ju_tu_j]=0, \quad k=1,2,\ldots,n,
\end{equation}
where $E$ is the energy density defined in \eqref{E}.

The fact that in the one-dimensional case there exists  a simple change of dependent variable
\begin{equation}\label{trans-u}
  u(t,x)=\mu(t)^{-1/2}v(t,x)
\end{equation}
respecting the $u$-dependence of  the nonlinear interaction $f(u)$ and transforming the damped equation \eqref{main} to the undamped form \eqref{undamped}, namely removing the damping term $a(t)$ was studied in Ref. \cite{AncoMarquezGarridoGandarias2024}. This happens  when $a(t)$ satisfies
\begin{equation}\label{a-ode}
\frac{\dot{a}}{2}+\frac{a^2}{4}+\frac{\kappa}{2}\int a(t)dt=\sigma_0=\text{const.},
\end{equation}
and $f(u)$, which has no explicit time dependence, is a logarithmic interaction
\begin{equation}\label{log-u}
  f(u)=(\sigma+\kappa \ln |u|)u.
\end{equation}
$f(u)$ is transformed to $g(v)=(\sigma-\sigma_0+\kappa \ln |v|)v$. This means, by transformation, the constant $\sigma$  shifts as $\sigma\to \sigma-\sigma_0$.
The same result holds  in the multidimensional case. In the case when $a(t)$ can be set to zero, all known conservation law results as discussed before  are available for the undamped PDE (see formulae \eqref{cons-Pk-comp}-\eqref{angular-comp}) corresponding to the new nonlinear interaction $g(v)$.

The general solution of \eqref{a-ode} is not elementary. It can be expressed in terms of Lambert function. The elementary solutions exist for instance when $\kappa=0$, but in this case $g(v)$ becomes linear.

We conclude with the observation that, for the  damping term $a(t)=m/t$ ($\mu=t^m$), frequently used here, the left side of \eqref{a-ode} is
$$m\left[ \frac{1}{4}(m-2)t^{-2}+\frac{\kappa}{2}\ln |t|\right], \quad m\ne 0$$
which can never be a constant unless $m=2$, $\kappa=0$ for which the logarithmic nonlinearity would not be present.


\end{document}